\documentclass{aa}
\usepackage{epsfig}
\usepackage{txfonts}
\usepackage{graphics}
\begin{document}

   \title{Revisiting the radial distribution of pulsars in the Galaxy}

   \author{I. Yusifov  \and   I. K\"u\c c\"uk}

   \institute{Department of Astronomy \& Space Sciences,
              Faculty of Arts \& Sciences,\\
              Erciyes University, Talas Yolu, 38039 Kayseri, TURKEY\\
              \email{yusifov@erciyes.edu.tr} \\
              \email{kucuk@erciyes.edu.tr}
             }
 
 \offprints{I. Yusifov,\\
          e-mail: {\tt yusifov@erciyes.edu.tr}}
             
    \date{Received xxxx / Accepted 31 March 2004}

\abstract{
The high sensitivity Parkes and Swinburne Multibeam pulsar surveys have nearly doubled the number of known pulsars and revealed many more distant pulsars with high dispersion measures. These data allow us to investigate in more detail the statistical parameters and distribution of pulsars, especially in the central regions of the Galaxy, which was almost impossible in previous low-frequency and less-sensitive surveys. To estimate the distances to pulsars we used the new NE2001 Galactic electron density model. We study the Galactic distribution of normal pulsars with 1400 MHz luminosities greater than 0.1 mJy kpc$^2$ refining the shape and parameters of the radial distribution of  pulsars. \\ 
\noindent The maximum galactocentric distribution of pulsars is located at 3.2$\pm$0.4 kpc and the scale-length of this distribution is 3.8$\pm$0.4 kpc for the assumed distances to the Galactic center $R_\odot=8.5$kpc. The surface density of pulsars near the Galactic center is equal to or slightly higher than that in the solar neighborhood.\\
\noindent For observable normal pulsars with luminosities $\ge0.1$ mJy kpc$^2$, we also re-estimate their local surface density and birth-rate: 41$\pm$5 pulsars kpc$^{-2}$ and 4.1$\pm$0.5 pulsars kpc$^{-2}$ Myr$^{-1}$ respectively. For the total number of potentially observable pulsars in the Galaxy, we obtain $(24\pm3)\times10^3$ and $(240\pm30)\times10^3$ before and after applying beaming correction according the Tauris \& Manchester (1998) beaming model. Within the limits of errors of estimations these results are in close agreement with the results of the previous studies of Lyne et al. (1998)  (hereafter LML98).\\ 
 The dependence of these results on the NE2001 model and recommendations for further improvement of electron density distribution are discussed.\\

        }

\titlerunning{Radial distribution of pulsars}
\authorrunning{Yusifov \& K\"u\c c\"uk}

\maketitle

\section{Introduction}

Since the discovery of pulsars there have been numerous statistical investigations of their parameters (see for example \cite{go70}, \cite{tm77}, \cite{lmt85}, \cite{gy84}, \cite{n87}, LML98, and references therein), such as the Galactocentric radial and height distribution, the luminosity function, the evolution of magnetic field and luminosity of pulsars, etc. These results are then used for the estimation of the birth rate of pulsars and for the modeling of Galactic structures.

Early surveys of pulsars were carried out around frequencies of 400 MHz. At low frequencies the sky background temperature dramatically increases in the direction of the Galactic center (GC), and also the interstellar medium (ISM) electron density rises. According to the sensitivity relation (\cite{D84}, see also relation (\ref{Smin}) below) both of these effects lead to the reduction of search sensitivity toward the GC, which in turn, leads to the reduced detection of pulsars around the GC. It makes it difficult to determine the precise radial distribution of pulsars in the Galaxy as well as the precise estimation of the number density of pulsars at the GC. So far in various statistical investigations, various approximations of radial variation of number density of pulsars, such as filled center or hollow center distributions, were used (see for example \cite{n87}, \cite{h97},  \cite{gob2001},  \cite{ss02}, and references therein). A preliminary study of central regions of the Galaxy, on the base of early high frequency surveys (\cite{cljm92} and \cite{jlm92}), was carried out by Johnston (1994). The conclusion was the  deficit of pulsars near the GC.

\begin{figure*}[t]
\begin{center}
\includegraphics[width=16cm,height=10cm] {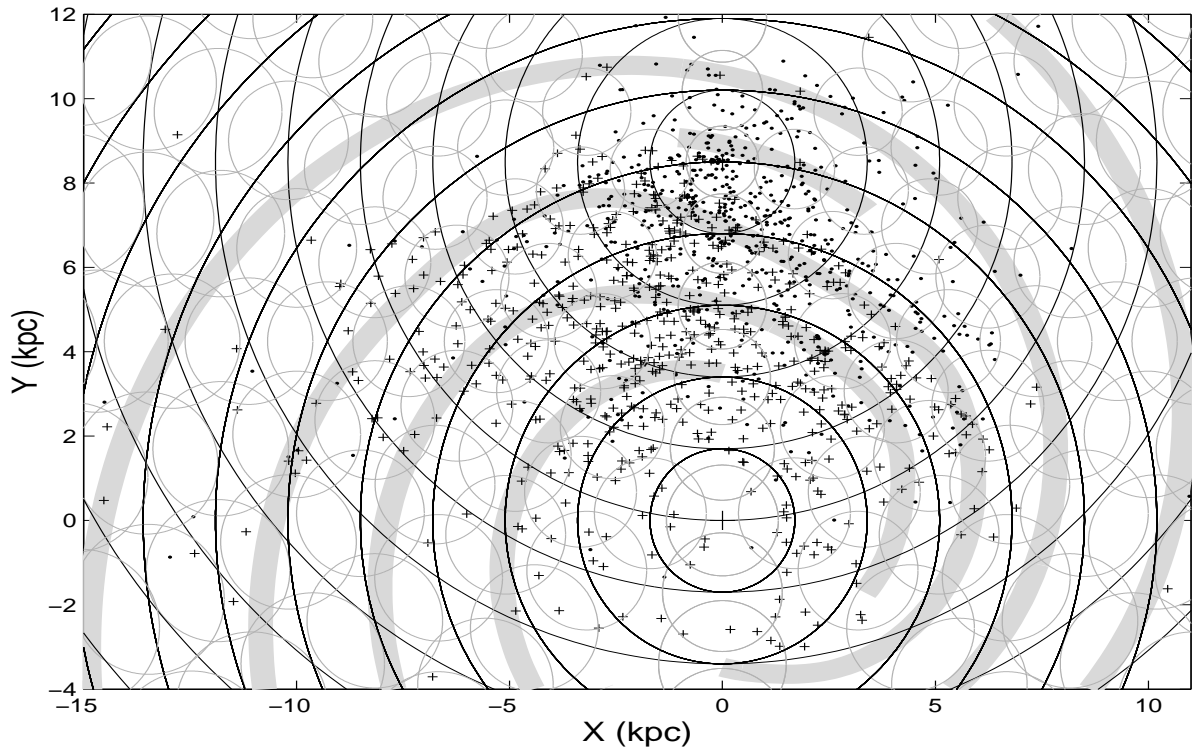}
\end{center}
\caption{Projection of pulsars onto the Galactic plane. Pulsars discovered in the MBPS survey are marked by pluses; pulsars known before are marked by dots. GC is located at the beginning of coordinate system. Spiral arms of the Galaxy are from CL2002. The (x, y) coordinates of the Sun are (0.0, 8.5). The function of other lines and symbols is explained in the text. }
\label{Fig:Galdistr} 
\end{figure*}

Recent high frequency, sensitive Parkes Multibeam Pulsar Surveys (PMPS) (\cite{mlc2001}, \cite{mhl2002} and \cite{kbm2003}) revealed many more distant pulsars with high Dispersion Measures (DM). These data may allow us to investigate in more detail the statistical parameters and distribution of pulsars, especially in the central regions of the Galaxy, which was almost impossible in previous low-frequency and less-sensitive surveys. Furthermore, high frequency searches reveal many pulsars at the lower as well as at the high luminosity ends of the luminosity function (LF) of pulsars (\cite{cam2002}, \cite{lor2003} ). From the evolutionary standpoint it is of great interest in understanding the influence of such pulsars on the shape of pulsars' LF. 

We decided to estimate the density of pulsars at the GC and to take a fresh look at the radial distribution and luminosity function of pulsars using these new PMPS data from the ATNF Pulsar Catalogue of 1412 pulsars (\cite{mancat}). We estimated the distances to the pulsars using the NE2001 Galactic electron density model, Cordes \& Lazio (2002) and Cordes \& Lazio (2003), (CL2002 and CL2003 hereafter). 

In the space and luminosity distribution studies, it is very important to make corrections for the observational selection effects. In various statistical studies, many authors (see for example \cite{lmt85}, \cite{NO90}, \cite{bwh92}, \cite{Lbd93}, \cite{j1994}, \cite{sd96}, \cite{CC97},
LML98, \cite{gob2001} and references therein)  applied various reliable but complicated statistical methods to precisely quantify important selection effects. In this paper we use an empirical (and simple) method for the correction of observational selection effects, as in Kodiara (1974), Yusifov (1981), Leahy \& Wu (1989) and Wu \& Leahy (1989) (WL89, hereafter), with some modifications. 

A summary of the available data and apparent distribution of pulsars in the Galaxy are presented in Section 2. In Section 3 we will discuss the selection effects and correction methods we applied. In Section 4 we derive the radial distribution of pulsars in the Galaxy. The discussion of the results and conclusions are presented in Sections 5 and 6.

\section{Available data and the apparent distribution of pulsars in the Galaxy }

At the time of preparing this paper, the number of existing pulsars was 1412 (\cite{mancat} ). The ATNF Pulsar Catalogue contains the data of PMPS, Swinburne (\cite{ebsb2001}) survey and all previous pulsar survey results, and it provides a good sample of data for the statistical study of pulsars. Nearly 600 of them were discovered in the PMPS survey. 

\begin{table*}[t]
\caption{The apparent density of pulsars}
\begin{tabular}{cccccccccccccc}
\hline
\begin{picture}(20,20)
      \put(15,2){R(kpc)}
      \put(-3,-13){r(kpc)}
      \thinlines
      \put(-5,20){\line(1,-1){35}}
   \end{picture}
     & &    0  &  1.7 &  3.4 &  5.1 &  6.8 &  8.5 & 10.2 & 11.9 & 13.6 & 15.3 &  17   & 18.7 \\
\\
\hline 
 0.0 & & $ $   &   $ $   &   $ $   &   $ $   &   $ $     &  38.0   &   $ $   &   $ $   &   $ $   &   $ $   &   $ $   &   $ $  \\
 1.7 & & $ $   &   $ $   &   $ $   &   $ $   &  31.16  &  16.12  &  5.73   &   $ $   &   $ $   &   $ $   &   $ $   &   $ $  \\
 3.4 & & $ $   &   $ $   &   $ $   &  17.52  &  15.14  &  11.88  &  4.06   &  0.99   &   $ $   &   $ $   &   $ $   &   $ $  \\
 5.1 & & $ $   &   $ $   &  10.5   &  11.63  &  7.25   &  6.75   &  2.35   &   0.6   &  0.04   &   $ $   &   $ $   &   $ $  \\
 6.8 & & $ $   &  3.84   &  8.22   &  6.51   &  5.87   &   4.7   &  1.69   &  0.26   &  0.07   &    0    &   $ $   &   $ $  \\
 8.5 & & 1.29  &  3.32   &  3.04   &  3.69   &  3.41   &  2.77   &  2.13   &  0.36   &   0.1   &  0.02   &    0    &   $ $  \\
10.2 & & $ $   &  1.45   &  1.69   &  1.13   &   0.8   &  1.05   &  1.27   &  0.66   &  0.09   &  0.03   &    0    &    0  \\
11.9 & & $ $   &   $ $   &  0.71   &  0.51   &  0.31   &  0.32   &   0.8   &  0.42   &  0.15   &  0.09   &  0.02   &    0  \\
13.6 & & $ $   &   $ $   &   $ $   &  0.18   &  0.11   &  0.06   &  0.23   &  0.26   &   0.2   &  0.07   &  0.07   &  0.01  \\
15.3 & & $ $   &   $ $   &   $ $   &   $ $   &  0.02   &  0.13   &  0.07   &  0.22   &  0.22   &  0.18   &  0.03   &  0.01  \\
17.0 & & $ $   &   $ $   &   $ $   &   $ $   &   $ $   &  0.06   &  0.08   &  0.04   &  0.12   &   0.2   &  0.05   &    0   \\
18.7 & & $ $   &   $ $   &   $ $   &   $ $   &   $ $   &   $ $   &  0.03   &  0.01   &  0.01   &  0.04   &  0.07   &  0.01  \\
\hline
\end{tabular}
\end{table*}

In this study we are interested mainly in statistics of "normal" pulsars. For this reason we excluded from our sample binary and recycled ($\dot{P} <10^{-17}$ s/s), globular cluster and Large and Small Magellanic Cloud pulsars. Of course, some pulsars in the remaining sample are related to the recycled ones, but at the  present stage it is difficult to separate them. This gives 1254 pulsars in total, and 1043 of them are located in the regions of Galactic longitudes $-100^\circ\leq l \leq 50^\circ$. In this sample, 581 pulsars were previously unknown and detected in the Parkes and Swinburne Multibeam Pulsar Surveys (MBPS). In our study we mainly used pulsars in the regions of the Galactic latitudes $|b|\leq 15^\circ$ and longitudes $-100^\circ\leq l \leq 50^\circ$. However, introducing the required corrections in density estimations, pulsars outside of this region were also taken into account, as described in the corresponding parts of the text.

Projection of these pulsars on the Galactic plane is shown in Fig.~\ref{Fig:Galdistr}. Although the recent observations (\cite{Eisen}) show that the distance to the GC is  8.0$\pm$0.4 kpc, in this study, as in the Galactic electron density model NE2001, we used the old value of the Sun-GC distance $R_\odot = 8.5$~kpc.

The apparent distribution of pulsars for the subsequent corrections due to selection effects is derived in the following manner. 
We drew equidistant concentric circles around the Sun ($r_i$) and the GC ($R_j$) and made up a quasi-regular grid of points at the points of intersection of these circles on the Galactic plane (see Fig.~\ref{Fig:Galdistr}). To subdivide the Sun-GC distance into a sufficient number of intervals, distances between adjacent circles were selected equal to  $\Delta R=R_\odot / 5=1.7$~kpc. Then we drew circles of radius $R_{ci}$ around the grid points ( or cells ) and counted the number of pulsars within the boundaries of circle $R_{ci}$. With the increase in distance from the Sun, the apparent density of pulsars decreases due to the selection effects. To avoid the loss of some pulsars at large distances from the Sun, the radius of the distant cells must be large. Arbitrarily, $R_{ci}$ increasing linearly is chosen from $0.1 R_\odot$ to $1/10$ of the largest considered distances from the Sun (18.7 kpc) and calculated  by the relation $R_{ci}=0.85*(1.2*(i-1)/11+1)$, where $i$ varies from 1 to 12. Large cell radii at larger distances naturally reduce the density fluctuations due to small number statistics and the selection effects. Grid circles around the Sun and GC were drawn until 18.7 kpc. Apparent densities of pulsars (in units 
pulsars~kpc$^{-2}$) in grid cells are shown in Table~1. 

Columns in Table~1 correspond to the cells at the same distances $R_j$ from the GC, which are indicated in the first row, and rows correspond to the cells at distances $r_i$ from the Sun, which are indicated in the first column.

Assuming galactocentric symmetry of distributions, the average densities in Table~1 are estimated in the following manner: for the regions $|l|\leq 50^\circ$  and $|l|\geq 100^\circ$, the number of pulsars in symmetric cells around the y axis is averaged ( i.e. for the cells with coordinates ( y, x ) and ( y, -x ) in Fig.~\ref{Fig:Galdistr}); for the region $-100^\circ\leq l \leq -50^\circ$, if the number of pulsars in the cells with positive x is more than in the negative x, we again averaged the numbers in the cells (y, x) and (y, -x). 

As is seen in Fig.~\ref{Fig:Galdistr}, at large distances from the Sun and the GC, the number of pulsars rapidly decreases, and in some cells pulsars are completely absent. In such cases, in order to reduce rapid fluctuations,in general, observational data 
are smoothed. Smoothing on the line of the Sun-GC direction, but behind  the GC and in the anticenter direction, is carried out by the relation: 

\begin{equation}
  \rho'_{ij} = (  \rho_{ij-1} +  \rho_{ij+1} + 4 \rho_{ij} +    
  \rho_{i-1j} +   \rho_{i+1j} ) / 8    
  \label{No2} 
,\end{equation}
where, $\rho'_{ij}$ is the smoothed surface density at the intersections of $i^{\rm th}$ and $j^{\rm th}$ circles correspondingly from the Sun and the GC.  But in other directions the situation is changed, and each cell is surrounded by 6 other cells. For this reason smoothing for the distances more than 10 kpc from the Sun and GC (corresponding to the columns $R_j>11$~kpc and rows $r_i>11$~kpc) is carried out by the relation:

\[
  \rho'_{ij} = ( \rho_{ij-1} +  \rho_{i-1j-1} +  \rho_{i-1j} + 6 \rho_{ij} +  
\]
\vspace{-5mm}
\begin{equation}
\hspace{4cm} \; \rho_{i+1j} +  \rho_{ij+1} +  \rho_{i+1j+1}) / 12, 
 \hfill  \label{No3} 
\end{equation}
where designations are as in Eq. (\ref{No2}). The densities in Table~1 are obtained after applying these averaging and smoothing of observational densities. In these procedures, smoothed values are not additionally re-smoothed. For this reason, the cells totally surrounded by the empty cells in Table 1 remain with zero density.   

\section{The selection effects and correction method}

Data from Table~1, after correcting for the selection effects, may be used for the calculation of the radial distribution of pulsars in the Galaxy. 

In previous population studies (LML98) the number and birth-rate were estimated for the pulsars of 400 MHz luminosities greater than 1 mJy kpc$^2$. If we take into account the relation  $S\sim\nu^\alpha$ for pulsars, where $S$ is observed flux, $\nu$ is frequency and $<\alpha>=-1.7$ (\cite{S2002}), the corresponding minimum luminosity at 1400 MHz  is $L_{\rm min 1400}\approx$0.1 mJy kpc$^2$. Within a radius of 1 kpc from the Sun there are nearly 90 pulsars and only 10 of  them have 
1400 MHz luminosities  $\le0.1$mJy kpc$^2$.  In addition, for 4 of them flux densities at 1400 MHz are not known and the approximate luminosities are estimated from the assumed mean spectral indices $<\alpha>=-1.7$, so that the population of low luminosity pulsars still remains uncertain due to small number statistics.  For this reason, and for simplicity of  comparison of similar parameters of early results we consider the sample of pulsars with 1400 MHz luminosities greater than $L_{\rm min 1400}$.

WL89 considered three types of selection effects, one of which is connected to the much more sensitive Arecibo survey (\cite{Stokes}) in comparison with other ones (Green 
Bank (\cite{Stokes}), Jodrell Bank (\cite{cl86}) and Molonglo (\cite{man78})). To get a homogeneous sample, they excluded Arecibo pulsars in galactic longitudes $42^\circ\leq l \leq 60^\circ$. As we  considered in general MBPS survey region of the Galaxy ($-100^\circ\leq l \leq 50^\circ$), we examine only two types of selection effects with some modifications. 

We divide selection effects into two categories and define them by the relation:

\begin{equation}
\rho (r,R,l(r,R)) = K(l)K(r)\rho _o (r,R,l(r,R))
  \label{RorR} 
,\end{equation}
where $l$ is the Galactic longitude; $r$ and $R$ distances from the Sun and the GC; $K(l)$ direction and $K(r)$ distance-dependent selection factors; $\rho _o (r,R,l(r,R))$ is the true and  $\rho (r,R,l(r,R))$ is the observed distribution of surface densities of pulsars on the Galactic plane. Table 1 is a numeric representation of  $\rho (r,R,l(r,R))$.   

\begin{figure}
	\begin{center}
	\mbox{\epsfig{file=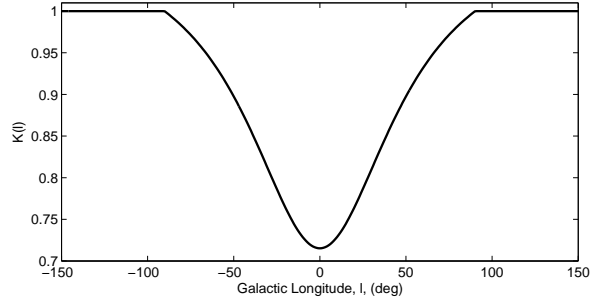,height=4cm}}
	\end{center}
	\caption{ Dependence of the correcting factor $K( l )$ as a function of Galactic longitude. }
	\label{Fig:K1(l)} 
\end{figure}

$K(l)$ is connected to the background radiation which leads to variations of the survey sensitivity with Galactic longitudes. Survey sensitivity is a function of several parameters, and in its simplest case may be described by the relation (see for example \cite{D84}, \cite{lmt85}): 

\begin{equation}
  S_{\rm min}=\beta S' \biggl(1+ {T_{\rm sky}\over T_{\rm R}} \biggr)
  \biggl(1+ {DM \over DM_0}\biggr) \quad {\rm mJy}
  \label{Smin} 
,\end{equation}
where $S_{\rm min}$ (mJy) is the minimum detectable flux; $S'$ (mJy) is the minimum detectable flux at the telescope beam center in the ideal situation; $\beta$ is a factor representing the reduction of sensitivity for the off-center scan of pulsars by a telescope beam;  $T_{\rm sky}$ is sky background temperature; $T_{\rm R}$ is receiver noise temperature; $DM$ is dispersion measure; and $DM_0$ is the characteristic dispersion measure depending on the antenna, receiver and other parameters of the survey. Sky temperature depends on observational frequencies and Galactic coordinates. The sky temperature map at 400~MHz has been determined by Haslam et al. (1982). Their results are analytically fitted by Narayan (1987) and have the form:

\begin{equation}
T_{\rm sky \, 400}(l,b)=25+
{275 \over [1+(l/42)^2][1+(b/3)^2]} \quad K
\label{Tsk400} 
,\end{equation}
where $l$ and $b$ are the Galactic longitude and latitude, respectively. This sky temperature, then, may be scaled to the sky temperature of the observing frequency $\nu$ (MHz) through the relation given by Johnston et al. (1992) in the form 

\begin{equation}
T_{\rm sky \, \nu}= T_{\rm sky \, 400} \biggl( {408{\rm MHz} \over \nu } \biggr)^{2.6} K
\label{Tsk1400} 
.\end{equation}
For the MBPS search frequencies we obtain: $T_{\rm sky \,1374}(l,b) = 0.04\,T_{\rm sky \,400}(l,b)$. As in WL89, the direction-dependent correcting factor $K(l)$ is defined by the relation: 

\begin{equation} 
K(l)=\biggl(1+ { T_{\rm sky\,1374}(l,b=0) \over T_{\rm R} }\biggr)^{-1}
\label{K1(l)} 
,\end{equation}
where $T_{\rm R}{=}21^\circ {\rm K}$ is the noise temperature of the Parkes multibeam receiver (\cite{mlc2001}). Neglecting latitude dependence may lead to some overestimation of observed densities. However, taking into account that the majority of distant pulsars from which radial distribution is derived are low-latitude pulsars, we may use Eq.~(\ref{K1(l)}) as the first approximation. The correction factor $K( l )$ which was calculated from Eqs. (\ref{Tsk400}$-$\ref{K1(l)}) is shown in Fig.~\ref{Fig:K1(l)}. We assume that reduction of sensitivity according to Eq. (\ref{Smin}), due to this selection effect, makes the observational number of pulsars in grid cells decrease. For this reason, the observational number (or density) must be corrected by this selection effect, dividing by $K( l )$. In the anticenter direction and in the cell where the Sun located ($r<0.85$~kpc), $K( l )$ is assumed to be 1.

Kodiara (1974) gives an empirical method for correcting distance-dependent selection effects for the study of SNRs. Further, in Huang et al. (1980), Yusifov (1981), Guseinov \& Yusifov (1984), Wu et al. (1984), Leahy \& Wu (1989), WL89 and Case \& Bhattacharya (1989) this method was used to study the Galactic distributions of pulsars and SNRs. With some modifications, the same method is applied here to study the Galactic distribution of pulsars from MBPS survey results.

In the flux-limited surveys for each distance ($d$) there is a minimum detectable luminosity defined by the relation

\begin{equation}
L_{\rm min}=S_{\rm min}d^2
\label{Lmin} 
,\end{equation}
where $S_{\rm min}$ is derived from Eq.~(\ref{Smin}) and, for the PMPS survey, is $\approx0.2$ mJy. Pulsars with luminosities lower than $L_{\rm min}$ are undetectable. Other reasons that reduce the sensitivity of pulsar surveys are the scattering and scintillation of pulsar radiation by the interstellar medium, pulse broadening, intrinsic variation of pulsar radiation, off center scan of pulsars, etc. 

\begin{table*}[t]

\caption{ Corrected densities of pulsars, in units pulsars kpc$^{-2}$}
\begin{flushleft}
\begin{tabular}{clcccccccccccc}
\hline
\begin{picture}(20,20)
      \put(15,2){R(kpc)}
      \put(-3,-13){r(kpc)}
      \thinlines
      \put(-5,20){\line(1,-1){35}}
   \end{picture}
     & &    0  &  1.7 &  3.4 &  5.1 &  6.8 &  8.5 & 10.2 & 11.9 & 13.6 & 15.3 &  17   & 18.7 \\
\\
\hline 
  0  & & $ $   &   $ $   &   $ $   &   $ $   &   $ $   &   41.5  &        $ $        &        $ $        &        $ $        &        $ $        &        $ $        &        $ $       \\
 1.7 & & $ $   &   $ $   &   $ $   &   $ $   &  80.62  &  30.12  &  {\bf  10.6  }    &        $ $        &        $ $        &        $ $        &        $ $        &        $ $       \\
 3.4 & & $ $   &   $ $   &   $ $   &  83.86  &  57.89  &  41.54  &  {\bf 13.89  }    &  {\bf  3.39  }    &        $ $        &        $ $        &        $ $        &        $ $       \\
 5.1 & & $ $   &   $ $   &  93.05  &   89.1  &  50.54  &  44.29  &       14.87       &  {\bf  3.81  }    &  {\bf  0.24  }    &        $ $        &        $ $        &        $ $       \\
 6.8 & & $ $   &  63.01  &  124.5  &  90.66  &  76.28  &  57.94  &       20.05       &  {\bf  3.11  }    &  {\bf  0.81  }    &  {\bf   0    }    &        $ $        &        $ $       \\
 8.5 & &39.16  &  98.28  &   84.9  &  96.05  &   83.5  &   64.4  &       47.74       &        7.92       &  {\bf  2.19  }    &  {\bf  0.33  }    &  {\bf   0    }    &        $ $       \\
10.2 & & $ $   &  81.29  &  89.65  &  55.85  &  37.45  &  46.19  &        53.7       &       27.08       &        3.48       &  {\bf  1.09  }    &  {\bf   0    }    &  {\bf   0    }   \\
11.9 & & $ $   &   $ $   &  73.55  &  48.92  &  27.57  &  26.79  &       64.66       &       32.65       &       10.96       &        6.48       &  {\bf  1.65  }    &  {\bf   0    }   \\
13.6 & & $ $   &   $ $   &   $ $   &  34.02  &  19.91  &  10.32  &       35.91       &       38.97       &       28.94       &        9.41       &       10.26       &  {\bf  1.44  }   \\
15.3 & & $ $   &   $ $   &   $ $   &   $ $   &   5.58  &   42.4  &       21.15       &       62.58       &       60.67       &       46.45       &        7.47       &        2.66      \\
 17  & & $ $   &   $ $   &   $ $   &   $ $   &   $ $   &   36.9  &       44.82       &       24.57       &        63.5       &         102       &        24.9       &         0        \\
18.7 & & $ $   &   $ $   &   $ $   &   $ $   &   $ $   &   $ $   &       30.67       &       11.41       &        8.46       &       39.34       &       64.99       &        6.77      \\
\hline
\end{tabular}
\end{flushleft}
\end{table*}

Pulse broadening, scattering and scintillation are strong functions of distances (see for example CL2003 and references therein). With increasing distances from the Sun each of these factors reduces the detected number (or density) of pulsars. Here we make the simplifying assumption that the combined effect of these factors define $K(r)$ in Eq.~(\ref{RorR}), which may be described by the exponential law as: 

\begin{equation}
K(r) = K_1 (r)K_2 (r) \cdots K_i (r) = e^{ - c_1 r} e^{ - c_2 r}  \cdots e^{ - c_i r}  = e^{ - cr} 
\label{K(r)} 
,\end{equation}
where $K_1(r)$, $K_2(r)$ etc. are the distance-dependent correction factors relating to the pulse broadening, scattering and other selection effects. The quantitative estimation of each of these factors independently is difficult, but the combined effect of these factors leading to reduced detection of pulsars away from the Sun may be estimated empirically as described below. 

We assume that the surface density (SD) of pulsars is symmetric around the GC, and considering a galactocentric circle with the radius $R_\odot$, from Eq.~(\ref{RorR}) we obtain: 
 
\begin{equation}
\rho (r,R_ \odot  ) = K(l)K(r)\rho _o (r,R_ \odot  )
\label{RorRo} 
,\end{equation}
where apparent densities $\rho (r,R_\odot)$ correspond to the $R=8.5$ kpc column in Table~1. According to the assumption $\rho _o (r,R_ \odot )$ is constant and $K(l)$ is known. 

\begin{figure} [!b]
\begin{center}
\mbox{\epsfig{file=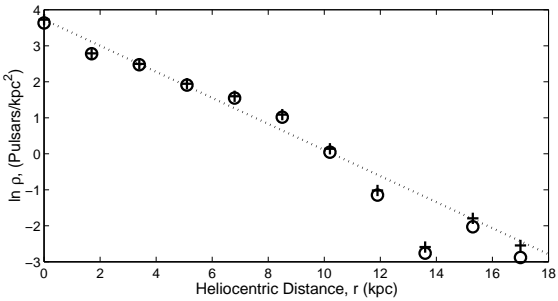,height=4cm}}
\end{center}
\caption{Variation of the surface densities of pulsars on the galactocentric circle $R=8.5$ kpc at various distances from the Sun (column 6 in Table~1). Apparent densities marked by circles and densities corrected due to the direction-dependent selection effect ($K(l)$ ) are marked by pluses. The dotted line is a LMS fitting of them}
\label{Fig:K2(r)} 
\end{figure}

The observed distribution of  $\rho(r,R_\odot)$ against the heliocentric distance $r$ is shown in Fig.~\ref{Fig:K2(r)}. We assume that observed exponential decay of SD in Fig.~\ref{Fig:K2(r)} is connected to the distance-dependent correction factor $K(r)$ in relations (\ref{RorR}), (\ref{K(r)}) and (\ref{RorRo}). Fitting the data in Fig.~\ref{Fig:K2(r)} with a simple exponent we can derive  $K(r)$ and $\rho _o(r,R_\odot)$.

The approximate  value of the local SD of pulsars $\rho _o(r,R_\odot)$ within $r \le 0.85$ kpc may be estimated using all available data, taking into account the different sensitivity of surveys as a function of the Galactic longitude and latitude. The method  of estimation is described in Yusifov \& K\"u\c c\"uk (2004). 
For the preliminary  value of $\rho _o(r,R_\odot)$ with 1400 MHz luminosities greater than $L_{\rm min 1400}$ we obtain 38 pulsar kpc$^{-2}$ marked by pluses in Fig.~\ref{Fig:K2(r)}.   
 A more precise value of $\rho _o(r,R_\odot)$ will be obtained extrapolating the fitting lines of distribution $\rho(r,R_\odot)$ 
 as a function of $r$ in Fig.~\ref{Fig:K2(r)}. 

Fitting the corrected data in Fig.~\ref{Fig:K2(r)} by the least mean squares (LMS) method for the apparent distribution of SD, we can derive $K(r)$ and $\rho _o(r,R_\odot)$ (during the fitting, the point around 13.6 kpc is not taken into account as an outlier):

\begin{equation}
\ln (\rho (r,R_ \odot )/K(l))=\ln (\rho _o K(r))=\ln \rho _o - cr
\label{Ror} 
,\end{equation}
where $ln\rho_o=3.73\pm0.16$ and $c=0.362\pm0.017$. The correction factor $K(r)$ we define as 

\begin{equation}
K(r)=\exp{(-cr)}
\label{K2(r)} 
,\end{equation}

The apparent densities ($\rho_{ij}$) in Table 1 are corrected by the relation  (\ref{RorR}) using K(l) and K(r) from Eqs.~ (\ref{K1(l)}) and (\ref{K2(r)}) and corrected SD  ($\rho_{ijc}$) of pulsars are given in Table~2. 

\begin{table*}[ht]

\caption{Errors of surface densities of pulsars estimated by the relation (12) }
\begin{flushleft}
\begin{tabular}{cccccccccccccc}
\hline
\begin{picture}(20,20)
      \put(15,2){R(kpc)}
      \put(-3,-13){r(kpc)}
      \thinlines
      \put(-5,20){\line(1,-1){35}}
   \end{picture}
     & &    0  &  1.7 &  3.4 &  5.1 &  6.8 &  8.5 & 10.2 & 11.9 & 13.6 & 15.3 &  17   & 18.7 \\
\\
\hline 
  0  & & $ $   &   $ $   &   $ $   &   $ $   &   $ $   &    4.8  &        $ $        &        $ $        &        $ $        &        $ $        &        $ $        &        $ $       \\
 1.7 & & $ $   &   $ $   &   $ $   &   $ $   &    9.5  &    4.7  &  {\bf   2.7  }    &        $ $        &        $ $        &        $ $        &        $ $        &        $ $       \\
 3.4 & & $ $   &   $ $   &   $ $   &   12.4  &    9.0  &    7.2  &  {\bf   3.9  }    &  {\bf   1.9  }    &        $ $        &        $ $        &        $ $        &        $ $       \\
 5.1 & & $ $   &   $ $   &   16.9  &   15.6  &   10.5  &    9.5  &         5.1       &  {\bf   2.5  }    &  {\bf   0.6  }    &        $ $        &        $ $        &        $ $       \\
 6.8 & & $ $   &   16.7  &   25.1  &   19.8  &   17.3  &   14.2  &         7.5       &  {\bf   2.8  }    &  {\bf   1.4  }    &  {\bf   0    }    &        $ $        &        $ $       \\
 8.5 & & 15.9  &   27.4  &   24.5  &   25.8  &   23.1  &   19.2  &        15.8       &         5.8       &  {\bf   3.0  }    &  {\bf   1.2  }    &  {\bf   0    }    &        $ $       \\
10.2 & & $ $   &   30.7  &   31.9  &   23.3  &   18.0  &   19.9  &        21.4       &        14.2       &         4.8       &  {\bf   2.7  }    &  {\bf   0    }    &  {\bf   0    }   \\
11.9 & & $ $   &   $ $   &   36.2  &   27.7  &   19.5  &   18.7  &        30.2       &        20.1       &        11.1       &         8.4       &  {\bf   4.2  }    &  {\bf   0    }   \\
13.6 & & $ $   &   $ $   &   $ $   &   29.7  &   21.5  &   14.7  &        27.7       &        28.4       &        23.8       &        13.0       &        13.5       &  {\bf   5.0  }   \\
15.3 & & $ $   &   $ $   &   $ $   &   $ $   &   15.0  &   40.7  &        27.2       &        47.5       &        45.8       &        38.9       &        14.8       &         8.7      \\
 17  & & $ $   &   $ $   &   $ $   &   $ $   &   $ $   &   50.6  &        53.3       &        37.5       &        60.5       &        77.5       &        35.6       &         0.0      \\
18.7 & & $ $   &   $ $   &   $ $   &   $ $   &   $ $   &   $ $   &        59.1       &        33.8       &        28.0       &        59.8       &        76.6       &        23.6      \\
\hline
\end{tabular}
\end{flushleft}
\end{table*}

Extrapolating the fitted line in Fig.~\ref{Fig:K2(r)}, for the local SD of pulsars with luminosities more than  0.1 mJy kpc$^2$, we obtain $\rho_0=41\pm5$ pulsars kpc$^{-2}$. Although the obtained local SD of pulsars slightly exceeds the value derived by LML98  (30$\pm$6), within the errors of estimations the values are in good agreement. If we take into account some possible reasons leading to an increase of the apparent local SD of pulsars, discussed in section 5, the agreement is further strengthened.

\section{Radial distribution of pulsars in the Galaxy}

By averaging the corrected SD of pulsars
in Table~2 we can derive the SD of pulsars
as a function of radial distance from the GC.

As the selection effects vary with the direction and distance of the Sun, errors of corrected densities also vary from point to point. The errors of corrected densities in corresponding cells in general were estimated  on the basis of Eq. (\ref{RorR}). 

We assume that errors in calculating corrected densities $\rho_{ijc}$ are mainly connected to the errors in the correction factor $K(r)$ and apparent density $\rho_{ij}=N_{ij}/\Delta S_i$, where $N_{ij}$ and $\Delta S_i$ are the number of pulsars in the cell and the area of the cell ($i,j$). $K(r)$ is an exponential function (Eq.~(\ref{K2(r)})) similar to $y=a \exp(bx)$, where the error due to uncertainty in $x$ is estimated with the standard relation (\cite{Bev69}): $\sigma_y/y = b \sigma_x$. In evaluating errors for $N_{ij}$ we assumed that the number of pulsars in cells, in a rough approximation, follows Poisson statistics, that is the mean value equals the square of the standard deviation ($\mu=\sigma^2$). Depending on these assumptions, errors for $\rho_{ijc}$ can be calculated by the relation

\begin{equation}
{\sigma_{\rho_{ijc}}^2 \over \rho_{ijc}^2} =
\biggl( { \sqrt{N_{ij}} \over N_{ij}} \biggr)^2 + (r\sigma_c)^2
\label{Sigmaro} 
,\end{equation}
where $\sigma_{\rho_{ijc}}$ is the error of $\rho_{ijc}$ ,  and $\sigma_c=0.017$ is the standard deviation of $c$ from Eq.~(\ref{Ror}).  In evaluating errors for the circumsolar region (where $r=0$ in Eq.~(\ref{Sigmaro})) we took into account the error for $\rho_0$ in Eq.~(\ref{Ror}). Errors of densities for each cell of Table 2 estimated by this method are given in Table 3. From Tables 2 and 3 we can estimate the mean value and errors of SD of pulsars. As the errors for various data points are not equal, we calculated the weighted average of SD and corresponding errors for every column (or corresponding distances from GC) of Tables 2 and 3 by the following equations:

\begin{equation}
\rho(R)= {\sum\limits_i {\rho_{ijc} \over \sigma_{ijc}^2} \over 
\sum\limits_i \sigma_{ijc}^2} \quad {\rm and} \quad 
\sigma^2(R)=\biggl( \sum_i(\sigma_{ijc}^{-2}) \biggr)^{-1}
\label{Rosigma} 
,\end{equation}
where the summation is  carried out by index $i$ on every column $j$. The radius $R_j$ from GC corresponding to column $j$ is given in the first rows of Tables $1-3$. During the calculation of $\rho(R)$ and $\sigma(R)$, the densities marked in bold in Tables~2 and 3 are not taken into account, since they remain outside the region  $-100^\circ\leq l \leq 50^\circ$. 

\begin{figure} [!b]
\begin{center}
\mbox{\epsfig{file=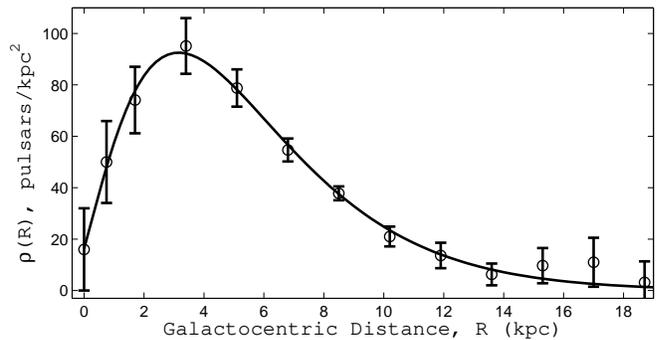,height=4.5cm}}
\end{center}
\caption{Radial distribution of surface densities of pulsars. Circles and error bars are derived from Eq.~(\ref{Rosigma}).
Fitting line corresponds to Eq.~(\ref{Gam14})}
\label{Fig:Ro} 
\end{figure}

\begin{figure}[!t]
\begin{center}
\mbox{\epsfig{file=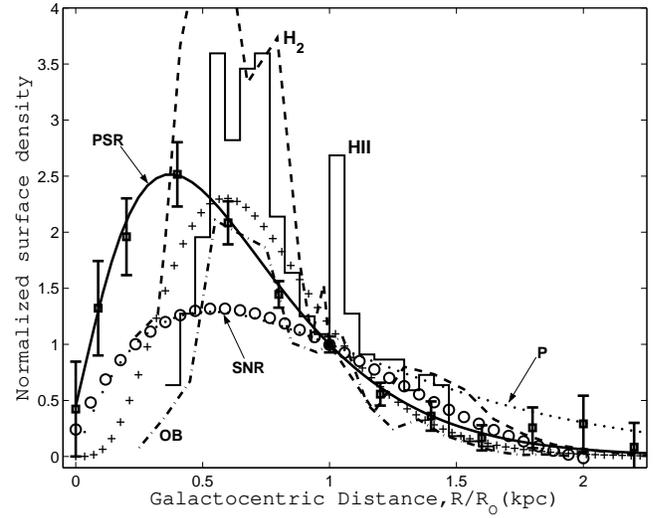,height=7cm}}
\end{center}
\caption{Radial distributions of pulsars (squares) and other types of Population I objects: SNR distributions are from Case \& Bhattacharya (1998); H2 column densities from Bronfman et al. (1988) and Wouterloot et al. (1990); HII regions are from Pladini et al. (2002), (HII regions represent the number of sources in the 0.5~kpc wide Galactocentric rings, but not surface densities); OB star formation regions data from Bronfman et al. (2000); the radial distributions of birth location of NS, from Paczynki (1990), (Eq.~(\ref{Pach1})) are  marked by {\bf P}; the expected radial distributions of birth location of NS (Eq.~(\ref{Pach2})) are marked by pluses }
\label{Fig:PopI}  
\end{figure}

From Table 2 we see that the SD of pulsars at the GC is
nearly $~40$ pulsar kpc$^{-2}$. This is the average density of pulsars within the cell of radius 1.3 kpc around the GC. In the considered sample, within the galactocentric radius of 0.5 kpc pulsars are absent, however, from Table 3 one can appreciate the error of the density estimation in this region as $\sim$16 pulsars kpc$^{-2}$. But within the 0.5 kpc $<$ R $<$ 1 kpc from the GC there are 4 pulsars. Applying correction factors $K(l)$ and $K(r)$ to the apparent density of pulsars around the galactocentric radius 0.75 kpc we obtain $~50$ pulsars kpc$^{-2}$. These values are shown in Figs.~\ref{Fig:Ro} and \ref{Fig:PopI} and are used in fitting the radial distribution. 

All the results of obtained radial distribution of SD of pulsars with corresponding error bars are plotted in Figs.~\ref{Fig:Ro} and \ref{Fig:PopI}. To simplify the comparison with other results, densities in Fig.~\ref{Fig:PopI} are normalized to the surface densities at the solar circle. Radial distribution of the SD in  Fig.~\ref{Fig:Ro} may be fitted by the frequently-used Gamma function written as:

\begin{equation}
\rho(R)=A\biggl({R \over R_\odot} \biggr)^a 
\exp{\biggl[-b\biggl({R-R_\odot \over R_\odot }\biggr) \biggr] }
\label{Gam14} 
,\end{equation}
where $R_\odot=8.5$ kpc is the Sun$-$GC distance. However, the 
relation (\ref{Gam14}) implies that  the surface density at $R=0$ is zero, which may be inconsistent with the reality. To obtain nonzero density at $R=0$ we include additional parameter $R_1$  and used a shifted Gamma function, replacing $R$ and $R_\odot$  in Eq.~(\ref{Gam14}) by $X=R+R_1$ and $X_\odot =R_\odot +R_1$ correspondingly. The best results of the fitting by the LMS method are: $ A=37.6\pm1.9 {\rm kpc}^{-2}, \> a=1.64\pm0.11$,  $b=4.01\pm0.24$ and $R_1=0.55\pm0.10$ kpc. 

The obtained radial distribution (Fig.~\ref{Fig:Ro}) shows a clear decrease of the SD of pulsars towards the GC, relative to the results of early studies at low frequencies (\cite{lmt85}).  The existence of pulsar deficit at the GC was already concluded by Johnston (1994), analyzing early high frequency pulsar surveys ( \cite{cljm92} and \cite{jlm92}).
MBPS allows to more precise determination of the density distribution of pulsars around the GC. As was shown earlier, at 0.75 kpc from the GC the SD of pulsars is around 50$\pm$16 pulsars kpc$^{-2}$  
and  is not less than that in the circumsolar region. Applying the Tauris \& Manchester (1998) (TM98 hereafter) beaming model, for this value we obtain (500$\pm$150) pulsars kpc$^{-2}$.  

At the present time there is a lot of information on radial distributions of various types of Galactic components, and it will be very interesting to compare radial distributions of pulsars and SNR with the radial distributions of their progenitors, shown in Fig.~\ref{Fig:PopI}.  

It is believed that, in general, NS are formed during SN II explosions and that their progenitors are OB type Population I stars. However, from Fig.~\ref{Fig:PopI} it is seen that, although the maximums of Population I objects coincide, the maximum of pulsar distribution is shifted towards the GC by nearly 1.5~kpc. The maximum of pulsar and SNR distributions nearly coincide, but the radial scale length ( RSL ) of the pulsar distribution ($\sim 4$~kpc) is nearly half that of the SNR distributions. 

The radial distribution of the birth location of NS is also striking. During population simulations of NS in the Galaxy, for the radial distributions of their birth location, the following relation derived by Paczynski (1990) (see for example \cite{sd96} and \cite{gob2001}) is used: 

\begin{equation}
\rho_R(R)=a_R\biggl({R \over R^{2}_{\rm exp}}\biggr)
\exp{\biggl( -{R \over R_{\rm exp}}\biggr)}
\label{Pach1} 
,\end{equation}
where $R_{\rm exp}=4.5$ kpc and $a_R=1.068$. The corresponding distribution is shown in Fig.~\ref{Fig:PopI} by dots and
marked as {\bf P}. Although the maximum of this distribution coincides with the maximum of the SNR distribution, its shape  considerably deviates from the distributions of Population I objects.
 
If the progenitors of  NS are the OB type Population I stars, then from the qualitative considerations of Fig.~ \ref{Fig:PopI}, it seems that the radial distribution of the  birth location of NS must be located near the line shown by pluses in Fig.~ \ref{Fig:PopI} and may be described by the relation: 
                   
\begin{equation}
\rho(R)=A\biggl({R \over R_\odot} \biggr)^a 
\exp{\biggl[-b\biggl({R \over R_\odot }\biggr) \biggr] }
\label{Pach2} 
,\end{equation}   
where $a=4$, $b=6.8$ and $R_\odot=8.5$ kpc is the Sun - GC distance. The constant A must be chosen from the calibration constraints (particularly, in the current case A = 1050). 
The final solution of  radial distributions of  birth locations of NS may be possible after performing a detailed analysis as in Paczynski (1990), taking into account the new data on pulsar velocities, radial distributions of pulsars and comparing obtained results with the new radial distribution data of Population I objects.

\section{Discussion}

From the derived distributions and age of pulsars one can estimate the total number of pulsars in the whole Galaxy, birthrate of pulsars and other statistical parameters. New age estimations of pulsars requires population synthesis or careful analysis of the $P-\dot{P}$ diagram and a new study of  $z$-distribution with more precise estimated distances of pulsars. However, from Kramer et al. (2003) and from our analysis it is obvious that distance estimations of pulsars require further improvement and additional studies. For these reasons, we estimated the birthrate of pulsars by assuming $\tau=10^7$yr as the mean lifetime of active pulsars, leaving detailed study to the future.

The total number of pulsars in the Galaxy may be estimated by integrating the radial distribution of SD from Fig.~\ref{Fig:Ro}. Integrating the relation (\ref{Gam14}) for the number of potentially observable pulsars with 1400 MHz luminosities greater than 0.1 mJy kpc$^2$, we obtain $(24\pm3)\times10^3$ and $(240\pm30)\times10^3$ before and after 
applying the mean value of the beaming factor (BF) from the TM98 model. For the corresponding  birthrates we obtain nearly one pulsar for every 400$\pm$50 and 40$\pm$5 yr in the Galaxy which is consistent with the LML98 results (one every 330 and 60 yr). Last (corrected) values are in close agreement with the recently derived rate of core collapsed supernovae (2 per century) in the Galaxy by \cite{McKeeW97} and (0.024-0.027)($H_o/75)^2$ yr$^{-1}$ by \cite{vandenberg}. In this relation $H_o$ is the Hubble constant.  

At the end of section 3 the local SD of pulsars was estimated as 41$\pm$5 pulsars kpc$^{-2}$. The corresponding local birth-rate of pulsars will be 4.1$\pm$0.5 pulsars kpc$^{-2}$ Myr$^{-1}$. While the new values of SD and birth-rate are higher, within the errors of estimations they are consistent with the LML98 results  (30$\pm$6 and 2.8$\pm$1.7 correspondingly).  

MBPS surveys revealed new pulsars not only in the central and distant regions of the Galaxy, but also some local faint pulsars in the circumsolar region. It seems likely that this is one of the reasons for the slightly high local density of pulsars obtained. Another reason for the apparent increase of the local SD probably is overestimated electron densities of ISM. This implies that the derived local SD is an upper limit and that the true value of the local SD of pulsars with luminosities greater than 0.1 mJy kpc$^2$ is located between 41 and 30 pulsars kpc$^{-2}$, in even better agreement  with the value derived by LML98 (30$\pm$6).

Applying beaming correction according to the TM98 model, for the local SD and birth-rate of pulsars we obtain $(410\pm50)$ pulsars kpc$^{-2}$ and 41$\pm$5 pulsars kpc$^{-2}$ Myr$^{-1}$, respectively. These values differ from the corresponding values of  LML98 (156 pulsars kpc$^{-2}$  and 10 pulsar kpc$^{-2}$ Myr$^{-1}$) by nearly 4 times. This is due to the BF used. We applied the TM98 BF ($\sim$10) which is nearly 2 times higher than the  \cite{Biggs} BF,  used in LML98.

In evaluating the statistical parameters of pulsars, the reliable estimation of distances is very important. Until recently the Taylor \& Cordes (1993) (TC93 hereafter) model has been widely used for this purpose. The PMPS survey reveals much more distant pulsars with high DM and the following estimations show that the distances for these pulsars calculated by the TC93 method are located far outside of the Galaxy (\cite{mlc2001}, \cite{mhl2002}, \cite{kbm2003}). 

The new electron density model developed by Cordes \& Lazio (2002) (NE2001) is free of these disadvantages, and there are no pulsars outside the Galaxy according to their model. Another advantage of their model is its ability  to be easily modified as further ISM data is collected (voids, clumps, HII regions).  

The mentioned electron density models consist of several components. One of them denoted as $n_1h_1$, the "thick disc" component and  due to its large scale-height ($h_1\approx 900$pc) it occupies a significant volume of the Galaxy. In the NE2001 model this value (0.033) is nearly two times higher than that of the TC93 model (0.0165). 

  Analyzing the NE2001 model, Kramer et al. (2003) discussed the  decreasing trends of mean $z$-height with increasing distances of pulsars from the Sun. But in the TC93 model the mean $z$-height increases away from the Sun. The most probable reasons for a similar variation of $z$-height may be related to the above mentioned 2 times higher value of $n_1h_1$ in the NE2001 model in comparison with the TC93 model. At the same time this also leads to some increase of local density of pulsars, as noted above.  Some intermediate value of $n_1h_1$  possibly may improve the variation of the mean $z$-height away from the Sun and the local SD of pulsars. However, this demands the fine tuning of the electron density distribution model in the Galaxy, which is beyond the scope of this paper.

In the future, in constructing new electron-density models, besides the improvements recommended by the authors of the NE2001, one must take into account the z-distribution of distant pulsars, new HI absorption measurements of distant pulsars and other statistical or integral parameters of the Galactic distributions of pulsars, derived in previous sections. Some improvements also may be achieved taking into account warp and flaring structure of the Galaxy (\cite{Y2003}). This may be possible after HI absorption measurements of a sufficient number of high DM pulsars. 

The dimensions of galaxies, RSL of stellar distribution, and the location of star formation regions are important parameters of galaxies. To estimate these parameters, in parallel with radio and optical data, researchers have applied infrared (IR) and other measurements of various galactic components.  For example, Drimmel \& Spergel (2001), by analyzing far-infrared (FIR) and near-infrared (NIR) radiation data from the COBE/DIRBE instrument, constructed a three dimensional model of the Galaxy and found the value $0.28R_\odot$ for the RSL of the stellar disc. In previous estimations this parameter has varied between 0.2 and 0.94 $R_\odot$(see \cite{Drim2001}, and references therein). 

OB stars are progenitors of NSs, and the radial distribution formation regions therein are studied in detail by Bronfman et al. (2000). The radial distribution of pulsars and OB star formation regions are compared in Fig.~\ref{Fig:PopI}, and despite the fact that the maxima do not coincide, the agreement is reasonable. In general, the majority of star formation regions and supernova (SN) activity are located in spiral arms and in the molecular ring around the GC. Recent investigations show that OB star-forming regions and SN activity may also be located at the periphery of the Galaxy (see \cite{Stil2001} and \cite{Bulent2002}). Distant pulsars with high DMs may play a crucial role in studying peripheral regions of the Galaxy. Once again this underlines the importance of reliable estimations of pulsar distances, especially for distant pulsars.     

\section{Conclusions}

We have studied the population of normal pulsars with the luminosities $L_{1400}\geq$ 0.1 mJy kpc$^2$ in the Galaxy, on the basis of the 1412 ATNF pulsar catalogue where the distances are calculated according to the new electron density model NE2001. Our results are as follows:

- We have refined the radial distribution of surface density of normal pulsars by using the new distances. The maximum of radial distribution is located at $3.2\pm0.4$~kpc and ${\sim}1.5$~kpc nearer to the GC relative to the maximum of distributions of Population I objects. Although the maximum of distribution nearly coincides with the maximum of distributions of their progenitors (SNR), the RSL of pulsars is  $3.8\pm0.4$~kpc, i.e. nearly two times less than the SNR distribution. Integrating the radial distribution for the total number of normal pulsars we obtained $(24\pm3)\times10^3$ and $(240\pm30)\times10^3$ before and after applying beaming corrections
according the TM98 model.

- The local surface density of pulsars is ($(41\pm5)$ and ($(410\pm50)$ pulsars kpc$^{-2}$) before and after beaming corrections.  

- The surface density of pulsars around the GC region (R$\approx$0.7 kpc) is (50$\pm$16) pulsars kpc$^{-2}$ and (500$\pm$150) pulsars kpc$^{-2}$ before and after applying beaming corrections. 

- The mean value of birth-rate of pulsars in the Galaxy is nearly one pulsar for every 400$\pm$50 and 40$\pm$5 yr before and after beaming corrections. 

- We recommend a new relation (Eq.~(\ref{Pach2})) for the expected radial distribution of birth location of NSs, which is more closely related to the radial distribution of Population I objects.

\begin{acknowledgements}

We would like to thank R.N. Manchester and the Parks Multibeam Pulsar Survey team for making the parameters of new pulsars available on the internet prior to formal publication as the ATNF catalogue of 1412 pulsars. We gratefully acknowledge the anonymous referee, for very detailed comments and suggestions, which led to significant improvements of the paper. We thank R. Wielebinski, J.L. Han and F.F. \"Ozeren for reading the manuscript and for useful discussions. We thank  Victor B. Cohen for help in preparing the manuscript. This work has been partially supported  by Erciyes University R/D project No. 01$-$052$-$1, Turkey. Extensive use was made of both the Los Alamos preprint archive and the ADS system.

\end{acknowledgements}

\noindent \textbf{Note added in proof:} 

\noindent {\footnotesize Recently, a paper of Lorimer, D.R., preprint[astro-ph/0308501] on the Galactic population of pulsars has been published. Although for study of pulsars they used different method, within the limits of the estimated errors, the obtained radial distribution and the total number of pulsars in the Galaxy (25$\pm$2)$\times10^3$  are nearly the same as in this paper.}

\end{document}